\documentclass
[prd,preprint,showpacs,groupedaddress]{revtex4-1}
\usepackage{amsmath}
\usepackage{amsfonts}
\usepackage{amssymb}
\usepackage{geometry}
\usepackage{natbib}
\usepackage[english]{babel}
\usepackage{graphicx}
\usepackage{epstopdf}
\begin{document}

  \title{Insight into the Microscopic Structure of an AdS Black Hole from the Quantization}

  \author{Dao-Quan Sun} \author{Jian-Bo Deng}  \email[Corresponding author:]{dengjb@lzu.edu.cn}, \author{Ping Li} \author{Xian-Ru Hu}

  \affiliation{Institute of Theoretical Physics, LanZhou University,
    Lanzhou 730000, P. R. China}

  \date{\today}

  \begin{abstract}

  	We explore the possible microscopic structure of a charged AdS black hole from the quantized viewpoint. A further study shows that some black holes cannot absorb ``energy quantum"  under certain conditions from the view of quantization.
 By the quantization of the black hole horizon area, we show the relation between the number of quanta of area and the microscopic degrees of freedom of the black hole. We also interpret a latent heat of thermodynamical phase transition as a transition between the number of quanta of area of large black hole(LBH) and the number of quanta of area of small black hole(SBH) in the charged AdS black hole. Furthermore, the Ruppeiner scalar curvature connecting with the number of quanta of area is also shown.

  \end{abstract}



  \maketitle

  \section{INTRODUCTION}

Over the past decades, black holes are widely considered as an important candidate to provide a bridge between a possible quantum theory of gravity and the classical general relativity. Black hole thermodynamics implies a fundamental relationship between gravitation, thermodynamics, and quantum theory~\cite{Hawking1975}. A pioneering work of black holes thermodynamics was done by Hawking and Bekenstein~\cite{PhysRevD.7.2333,Bekenstein1972,Hawking1975} who disclosed that temperature and entropy of a black hole on the event horizon satisfy the first law of thermodynamics. Moreover thermodynamic properties of anti-de Sitter (AdS) black holes have been of great interest since the Hawking and Page's seminal paper~\cite{Hawking1983} demonstrated the existence of a certain phase transition that can be in
stable equilibrium at a fixed temperature in the phase space of Schwarzschild-AdS black hole.
Since then the black hole phase transition and critical phenomena have been extended to a variety of more complicated backgrounds~\cite{1126-6708-1999-04-024,HENNEAUX1984415,HENNEAUX1984415,TEITELBOIM1985293}. If the cosmological constant was treated as the thermodynamic pressure~\cite{0264-9381-28-12-125020,kubizvnak2012p,Kastor:2009wy}, its conjugate quantity could be treated as  thermodynamic volume of the AdS black hole.
In terms of the pressure and volume, the charged AdS black hole shows small-large black hole thermodynamic phase transitions which is identified with liquid-gas phase transitions of the van der Waals(vdW) fluid~\cite
{kubizvnak2012p}.

	A black hole can change its Hawking temperature by absorbing or emitting matter, this can conjecture that black holes should have a microscopic structure, even though we do not know its micromolecules. Ruppeiner thermodynamic geometry~\cite{ruppeiner1995riemannian,ruppeiner2014thermodynamic,ruppeiner2008thermodynamic} gives phenomenologically a potential description about the types of interaction among micromolecules both in an ordinary thermodynamic system and in a black hole system.
In the Letter~\cite{wei2015insight}, authors introduced important concept of the number density of black hole micromolecules, and explored the possible microscopic structure of a charged AdS black hole completely from the thermodynamic viewpoint.

In 1974, Bekenstein~\cite{bekenstein1974quantum} presented that the black hole area should be represented by a quantum operator with a discrete spectrum of eigenvalues. By proving that the black hole horizon area is an adiabatic invariant, Bekenstein gave the form of the horizon area spectrum of black hole which is~\cite{bekenstein1973black,bekenstein1974quantum,hod1998bohr}
$A_{n}=\Delta A \cdot n$, where $\Delta A$ is the quantum of black hole area and Bekenstein gave by himself that $\Delta A = 8\pi l_{P}^{2}$. In addition, Hod~\cite{hod1998bohr} found a similar lower bound and gave $\Delta A = 4l_{P}^{2}$. He also suggested~\cite{hod1998bohr} that the black hole's area quantization can be determined by utilizing the quasi-normal modes frequencies of an oscillating black hole.  Since then, the subject of quantization of black hole has been extended by researchers.

In this paper, we explore the possible microscopic structure of a charged AdS black hole from the black hole area quantization to research the relation between the number of quanta of area and the microscopic degrees of freedom of the black hole. We use the quantization viewpoint of the black hole to interpret a latent heat of thermodynamical phase transition in the charged AdS black hole. Moreover, the Ruppeiner scalar curvature connecting with the number of quanta of area is also shown.

 The paper is organized as following. In Section \uppercase\expandafter{\romannumeral2}, we recall AdS black hole background of thermodynamics and quantization. In Section \uppercase\expandafter{\romannumeral3}, we study a jump between the two energy levels for black holes. The microscopic structure of a charged AdS black hole from the black hole area quantization is studied in Section \uppercase\expandafter{\romannumeral4}. Section \uppercase\expandafter{\romannumeral5} is reserved for conclusions and discussions.

\section{THERMODYNAMICS AND QUANTIZATION REVIEW in charged ADS BLACK HOLE}

In this section, we would like to review some basic thermodynamic properties of the spherically symmetric charged AdS black hole. For a four-dimensional charged AdS
black hole, using geometry units $\hbar=G=c=k_{B}=1$, the analogy between the vdW fluid and black hole system, authors~\cite{kubizvnak2012p} gave the equation of state
\begin{equation}
P=\frac{T}{2r_{h}}-\frac{1}{8\pi r_{h}^{2}}+\frac{Q^{2}}{8\pi r_{h}^{4}},
\end{equation}
where $r_{h}$ is the horizon radius, $Q$ is the total charge and $T$ represents temperature of the black hole. The charged AdS black hole was established by treating the cosmological constant as a pressure, $P=- \frac{\Lambda}{8\pi}$~\cite{0264-9381-28-12-125020,kubizvnak2012p}.
Comparing with the vdW equation of state, Ref.~\cite{kubizvnak2012p} identified the specific volume as
\begin{equation}
v=2l_{p}^{2} r_{h},
\end{equation}
where $l_{p}$ is the Planck length, $l_{p}=\sqrt{\frac{\hbar G}{c^{3}}}$. This concept has been a great success in studying small-large black hole phase transition. The critical point is obtained from
\begin{equation}
\frac{\partial P}{\partial v}=0     \\   ,\  \  \\   \frac{\partial P^{2}}{\partial v^{2}}=0,
\end{equation}
one can get
\begin{equation}
\label{eq:ljd-c}
T_{c}=\frac{\sqrt{6}}{18\pi Q},\  v_{c}=2\sqrt{6}Q,\  P_{c}=\frac{1}{96\pi Q^{2}}.
\end{equation}
In the paper~\cite{wei2015insight}, an important concept that is the number density of
black hole molecules as a measure for microscopic degrees of freedom of the black hole is introduced by
\begin{equation}
\label{eq:md-1}
\rho=\frac{1}{v}=\frac{1}{2l_{p}^{2} r_{h}},
\end{equation}
which could be naturally given an interpretation from the view of holographic in~\cite{wei2015insight,ruppeiner2008thermodynamic}.
Ruppeiner~\cite{ruppeiner2008thermodynamic} proposed that the microscopic degrees of freedom of the black hole are carried by the Planck area pixels. By assuming that one microscopic degree of freedom occupies $\gamma$ Planck area pixels, the total number of the microscopic degrees of freedom is
\begin{equation}
\label{eq:wgzyd-1}
N=\frac{A}{\gamma l_{p}^{2}},
\end{equation}
and the number density of the charged AdS black
hole can be calculated as~\cite{altamirano2014thermodynamics}
\begin{equation}
\label{eq:eq:md2}
\rho=\frac{N}{V}=\frac{3}{\gamma l_{p}^{2}r_{h}},
\end{equation}
where $V$ is the thermodynamic volume. Taking $\gamma=6$, it will obtain Eq.\eqref{eq:md-1}.

From the quantization of the black hole horizon area viewpoint, the black hole horizon area may be written~\cite{hod1998bohr}
\begin{equation}
\label{eq:mjlzh-1}
A_{n}=\alpha l_{p}^{2} \cdot n \   \ \ ; \  \   \  n=1, 2, \dots \ ,
\end{equation}
where $\alpha$ is a dimensionless constant.

\section{A POSSIBLE EXCHANGE OF ENERGY BETWEEN THE TWO STATES OF BLACK HOLES}
In the above section, we have reviewed the black hole area quantization. In this section we will study the behavior of a jump between the two energy levels for a black hole.
The metric of a nonrotating charged AdS black hole can be written as
\begin{equation}
ds^2=-f(r)dt^2+\frac{dr^2}{f(r)}+r^2d\Omega^2,
\end{equation}
with
\begin{equation}
f(r)=1-\frac{2M}{r}+\frac{r^2}{R^2}+\frac{Q^2}{r^2},
\end{equation}
where $R$ is the AdS radius, $d\Omega^2=d\theta^{2}+sin^{2}d\phi^{2}$, and the parameter $M$ indicates the ADM mass of the black hole solution while its horizon radius, $r_h$, is determined by the largest real root of $f(r)=0$ giving
\begin{equation}
\label{eq:m-r}
M=\frac{r_{h}}{2}+\frac{r_{h}^3}{2R^2}+\frac{Q^2}{2r_{h}}.
\end{equation}
According to the event horizon area
\begin{equation}
A=4\pi r_{h}^{2},
\end{equation}
the mass of the charged AdS black hole can be written as
\begin{equation}
\label{eq:m-A}
M=\frac{\sqrt{\frac{A}{\pi}}}{4}+\frac{(\frac{A}{\pi})^{\frac{3}{2}}}{16R^2}+\frac{Q^2}{\sqrt{\frac{A}{\pi}}}.
\end{equation}
From Eqs.\eqref{eq:m-A} and \eqref{eq:mjlzh-1}, the quantized mass of the charged AdS black hole can be written as
\begin{equation}
\label{eq:Mn-A}
M_{n}=\frac{\sqrt{\frac{\alpha m_{p}^{2}n}{\pi}}}{4}+\frac{(\frac{\alpha m_{p}^{2}n}{\pi})^{\frac{3}{2}}}{16R^2}+\frac{Q^2}{\sqrt{\frac{\alpha m_{p}^{2}n}{\pi}}}.
\end{equation}
Where $m_{p}$ is Planck mass, $m_{p}=\sqrt{\frac{\hbar c}{G}}$. In geometrized units G=c=1 and $l_{p}=m_{p}$.
Thus, for the charged AdS black hole, the energy difference between the two levels is written as
\begin{equation}
\label{eq:En-m}
\Delta E_{n_{1}\to n_{2}}= |E_{n_{2}}-E_{n_{1}}|=|M_{n_{2}}-M_{n_{1}}|,
\end{equation}
where $E_{n_{1}\to n_{2}}$ is interpreted as a jump between the two levels due to the energy be emitted or absorbed by black hole at that time.
One can compute an energy jump between two neighboring levels $n+1$ and $n$ is
\begin{equation}
\label{eq:En-1-n}
\begin{aligned}
\Delta E_{n \to n+1}&=E_{n+1}-E_{n}\\
&=\frac{\sqrt{\frac{\alpha m_{p}^{2}}{\pi}}}{4}(\sqrt{n+1}-\sqrt{n})+\frac{(\frac{\alpha m_{p}^{2}}{\pi})^{\frac{3}{2}}}{16R^2}[(n+1)^{\frac{3}{2}}-(n)^{\frac{3}{2}}]+\frac{Q^2}{\sqrt{\frac{\alpha m_{p}^{2}}{\pi}}}(\frac{1}{\sqrt{n+1}}-\frac{1}{\sqrt{n}}).
\end{aligned}
\end{equation}
The prime derivative of $\Delta E_{n \to n+1}$ with repect to $n$, one gets
\begin{equation}
\label{eq:dE-n}
\frac{d\Delta E_{n \to n+1}}{dn}=\frac{\sqrt{\frac{\alpha m_{p}^{2}}{\pi}}}{8}(\frac{1}{\sqrt{n+1}}-\frac{1}{\sqrt{n}})+\frac{3(\frac{\alpha m_{p}^{2}}{\pi})^{\frac{3}{2}}}{32R^2}(\sqrt{n+1}-\sqrt{n})+\frac{Q^2}{2\sqrt{\frac{\alpha m_{p}^{2}}{\pi}}}[(n)^{-\frac{3}{2}}-(n+1)^{-\frac{3}{2}}].
\end{equation}

\begin{figure}[htbp]
\centering
\includegraphics[width=7.2cm]{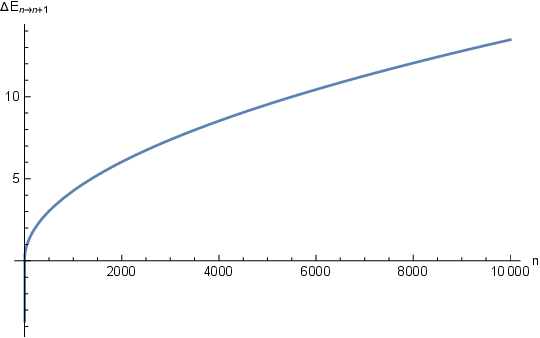}
\includegraphics[width=7.3cm]{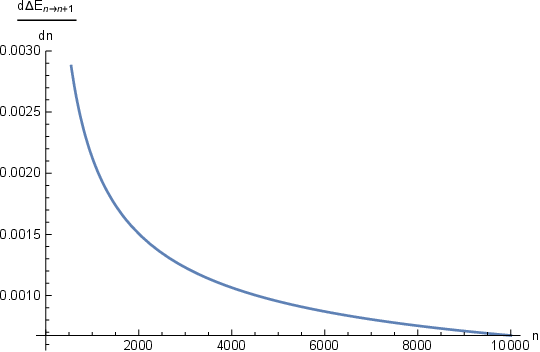}
\caption{
The figures depict the qualitative behavior of a jump between the two energy levels for a charged AdS black hole. Left:~$\Delta E_{n \to n+1}$-n plane. Right:~$\frac{d\Delta E_{n \to n+1}}{dn}$-n plane. We have set \(\alpha=4, m_{p}=1, Q=1, R=1\).}
\label{fig:1}
\end{figure}
From the Fig.~\ref{fig:1}, we can see that the sign of $\frac{d\Delta E_{n \to n+1}}{dn}$ is always positive.  One can also see that $\Delta E_{n \to n+1}$  always increases as $n$ increases, when $n=1$, $\Delta E_{n \to n+1}$ takes the minimum value $(\Delta E_{1 \to 2})_{min}$.
So, when an ``energy quantum" passes through the AdS black hole of $n\gg 1$, if this ``energy quantum" is less than the value of $\Delta E_{n \to n+1}$, it will not be able to be absorbed by this black hole.

In the same way, we can also discuss other black holes, such as Schwarzschild black holes, we can compute an energy jump between two neighboring levels $n$ and $n+1$. One obtains
\begin{equation}
\Delta E_{n \to n+1}=\frac{\sqrt{\frac{\alpha m_{p}^{2}}{\pi}}}{4}(\sqrt{n+1}-\sqrt{n}).
\end{equation}
The prime derivative of $\Delta E_{n \to n+1}$ with respect to $n$, one gets
\begin{equation}
\label{eq:SWXEn-1-n}
\frac{d\Delta E_{n \to n+1}}{dn}=\frac{\sqrt{\frac{\alpha m_{p}^{2}}{\pi}}}{8}(\frac{1}{\sqrt{n+1}}-\frac{1}{\sqrt{n}}).
\end{equation}

\begin{figure}[htbp]
\centering
\includegraphics[width=10.0cm]{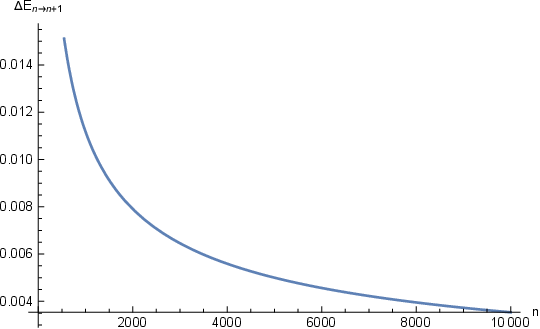}
\caption{
The figures of ~$\Delta E_{n \to n+1}$-n plane depict the qualitative behavior of a jump between the two energy levels for a Schwarzschild black hole.  We have set \(\alpha=8\pi, m_{p}=1\).}
\label{fig:2}
\end{figure}

From the Fig.~\ref{fig:2}, analysing the Eq.\eqref{eq:SWXEn-1-n}, we know when $n=1$, $\Delta E_{n \to n+1}$ takes the maximum value $(\Delta E_{1 \to 2})_{max}=1-\frac{\sqrt{2}}{2} \approx 0.29$.
Thus, when an ``energy quantum" passes through the Schwarzschild black hole of $n=1$, if this ``energy quantum" is less than the value of $\Delta E_{1 \to 2}$, it will not be able to be absorbed by this black hole.
For $n \to \infty$, namely, on a macro scale, $\Delta E_{n \to n+1}\to 0$, the energy spectrum of Schwarzschild black holes becomes approximately continuous which shows that an ``energy quantum" passes through the Schwarzschild black hole of $n\gg 1$, it will be able to be absorbed by this black hole.

For a R-N black hole, we can compute an energy jump between two neighboring levels $n$ and $n+1$. One obtains
\begin{equation}
\label{eq:R-NEn-n+1}
\Delta E_{n \to n+1}=\frac{\sqrt{\frac{\alpha m_{p}^{2}}{\pi}}}{4}(\sqrt{n+1}-\sqrt{n})+\frac{Q^2}{\sqrt{\frac{\alpha m_{p}^{2}}{\pi}}}(\frac{1}{\sqrt{n+1}}-\frac{1}{\sqrt{n}}).
\end{equation}

\begin{figure}[htbp]
\centering
\includegraphics[width=10.0cm]{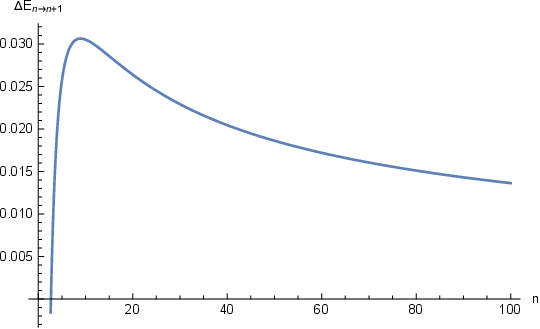}
\caption{
The figures of ~$\Delta E_{n \to n+1}$-n plane depict the qualitative behavior of a jump between the two energy levels for a R-N black hole. We have set \(\alpha=4, m_{p}=1, Q=1\).}
\label{fig:3}
\end{figure}

From the Fig.~\ref{fig:3}, analysing the Eq.\eqref{eq:R-NEn-n+1}, we know when $n=5$, $\Delta E_{n \to n+1}$ takes the maximum value $(\Delta E_{5 \to 6})_{max}$.
Thus, when an ``energy quantum" passes through the R-N black hole of $n=5$, if this ``energy quantum" is less than the value of $\Delta E_{5 \to 6})$, it will not be able to be absorbed by this black hole. For $n \to \infty$, namely, on a macro scale, $\Delta E_{n \to n+1}\to 0$, the energy spectrum of R-N black holes becomes approximately continuous, which shows that an ``energy quantum" passes through the R-N black hole of $n\gg 1$, it will be able to be absorbed by this black hole.

\section{ Insight into the Microscopic Structure of an AdS Black Hole from the Quantization}
In the previous section, we have reviewed some concepts of black hole molecules, microscopic degrees of freedom of the black hole, and discussed the behavior of the energy exchange between the two states of a black hole. In this section, we will discuss microscopic structure of a charged AdS black hole from the black hole area quantization.
Because Eq.\eqref{eq:wgzyd-1} implies the view of black hole thermodynamics, while Eq.\eqref{eq:mjlzh-1} implies the view of black hole quantization. One can compile these two ideas together then get
\begin{equation}
\label{eq:jg-1}
N=\frac{\alpha}{\gamma}n.
\end{equation}
From Eq.\eqref{eq:jg-1}, we establish the relationship between black hole area quantization and the microscopic degrees of freedom of the black hole. Namely, it establishes a relationship between the number of black holes molecules and the energy level of black hole. Thus, we can understand the thermodynamic molecular hypothesis of black hole from the perspective of energy level, which can advance our knowledge of about black holes thermodynamics.
This relationship also shows that the spectra of the microscopic degrees of freedom of the black hole are equidistant and dependent on the number of quanta of area.

If we take $n=1$ and $N\ge 1$, thus $\frac{\alpha}{\gamma} \ge 1$. Bekenstein~\cite{bekenstein1973black,bekenstein1974quantum,hod1998bohr} has given that the
assimilation of a finite size neutral particle causes to increase minimum quantum of black hole area that is $(\Delta A)_{min} = 8\pi l_{P}^{2}$. We can see $\alpha= 8\pi>\gamma=6$, thus if we require $N\ge 1$, then a Schwarzschild-AdS black hole can be in the ground state.
Hod~\cite{hod1998bohr} gave that the lower bound on the area increase caused by the assimilation of a charged particle is $(\Delta A)_{min} = 4 l_{P}^{2}$. We can see $\alpha= 4<\gamma=6$, thus, if we require $N\ge 1$, then the number of quanta of area $n$ must be $n\ge 2$, the charged AdS black hole
can not be in the ground state. Thus if we treat that black hole molecules as the real particles, only for the excited state of a charged AdS black hole, it may occur for the small-large black hole thermodynamic phase transitions.

Recalling a black hole system relationship~\cite{PhysRevD.7.2333,Bekenstein1972,Hawking1975} between horizon area
and its associated entropy, $S=\frac{k_{B}c^{3}A}{4\hbar G}$. Eq.\eqref{eq:eq:md2} may be written
\begin{equation}
\label{eq:lzhmd-1}
\rho=\frac{N}{V}=\frac{3}{\gamma l_{p}^{3}\sqrt{\frac{\alpha n}{4\pi}}}.
\end{equation}
The behaviors of the reduced number densities have been depicted in~\cite{wei2015insight,zangeneh2016comment,miao2018thermal}, when SBH crosses the coexistence curve and becomes a LBH.
Obviously, from Eq.\eqref{eq:lzhmd-1} we may also conjecture the behavior of the different the number of quanta of area between the SBH and LBH along the coexistence curve. With an increase of temperature $T/T_{c}$, the number of quanta of area $n/n_{c}$ discretely changes. The small-large black hole phase transition vanishes when approaching the critical point, which implies that the microscopic structures of the SBH and LBH tend to be the same, namely, there exists an unique state at that point. Here is the critical number of quanta of area $n_{c}=\frac{24\pi Q^{2}}{\alpha}$ got by Eqs.\eqref{eq:ljd-c} and \eqref{eq:lzhmd-1} with $\gamma=6$. If we take $\alpha=4$~\cite{hod1998bohr}, and then $n_{c}=6\pi Q^{2}$. So when the number of quanta of area $n$ must meet $n_{c}>n\ge 2$, we may conjecture that a small-large black hole phase transition occurs for a charged AdS
black hole.

For the black hole at fixed charge, the
latent heat $L$ of each black hole molecule transiting from one phase to another phase can be calculated from the following formula
\begin{equation}
\label{eq:qrgs}
\L=\frac{T\Delta S}{N}=\frac{T\gamma l_{p}^{3}}{3}\sqrt{\frac{\alpha}{4\pi}}(\sqrt{n_{LBH}}-\sqrt{n_{SBH}})\frac{dP}{dT}.
\end{equation}
From Eq.\eqref{eq:qrgs}, we can see the Clapeyron equation $dP/dT=\Delta S/\Delta V$ holding along the coexistence curve has been used, where the latent heat $L$ may be interpreted as a transition between the number of quanta of area of LBH and the number of quanta of area of SBH. While the latent heat vanishes when the system passes the critical point $n_{LBH}=n_{SBH}$.

The kind of intermolecular interaction along the transition curve for small and large black holes can be described by using the Ruppeiner geometry obtained from the thermodynamic fluctuation theory~\cite{ruppeiner1995riemannian,ruppeiner2014thermodynamic,ruppeiner2008thermodynamic}.
Ruppeiner thermodynamic geometry can tell us that the sign of $R$ determines the kind of
intermolecular interaction for the thermodynamic system
~\cite{ruppeiner1995riemannian,ruppeiner2014thermodynamic,ruppeiner2008thermodynamic}.
A positive thermodynamic scalar curvature $R$ corresponds to a repulsive interaction.
A negative thermodynamic scalar curvature $R$ shows an attractive interaction.
While thermodynamic scalar curvature $R=0$ indicates there is no interaction as in the case of classical ideal gas.
Now, we explore microscopic properties of the charged AdS black holes by applying thermodynamic geometry. In the paper~\cite{zangeneh2016comment}, the Ruppeiner geometry defined in $(M, P)$ space by taking entropy $S$ as thermodynamic potential while with fixed charge $Q=1$, authors figured out some microscopic properties of the 4-dimensional charged AdS black hole. The Ruppeiner scalar curvature is given by~\cite{zangeneh2016comment}
\begin{equation}
\label{eq:blzl-1}
R=\frac{1}{3\pi}\frac{(\rho / \rho_{c})^{6}-3(\rho / \rho_{c})^{4}}{-(\rho / \rho_{c})^{4}+6(\rho / \rho_{c})^{2}+(P/P_{c})}.
\end{equation}
While we got another form of Ruppeiner scalar curvature with simple calculation by Eqs.\eqref{eq:lzhmd-1} and \eqref{eq:blzl-1}
\begin{equation}
\label{eq:blzl--2}
R=\frac{1}{3\pi}\frac{(n_{c}/n)^{3}-3(n_{c}/n)^{2}}{-(n_{c}/n)^{2}+6(n_{c}/n)+(P/P_{c})}.
\end{equation}
When the denominator of Eq.\eqref{eq:blzl--2} is not zero and thermodynamic scalar curvature $R=0$, we may get $n=\frac{n_{c}}{3}$, it implies  no interaction of black hole micromolecules. When thermodynamic scalar curvature $R>0$, we may get $\frac{n_{c}}{3+\sqrt{9+P/P_{c}}}<n<\frac{n_{c}}{3}$, it corresponds to a repulsive interaction of black hole micromolecules. While thermodynamic scalar curvature $R<0$, we may get $n>\frac{n_{c}}{3}$ or $\frac{n_{c}}{3+\sqrt{9+P/P_{c}}}>n>1$, it corresponds to an attract interaction of black hole micromolecules.

\begin{equation}
\Lambda^\mu_\nu=\delta^\mu_\nu+\omega_{\rho \nu}\eta^{\rho \mu}.....\\ 
\rho \nu
\end{equation}

\section{conclusion}

In this paper, the possible microscopic structure of a charged AdS black hole from the quantization viewpoint has been explored. The possible energy exchange between the two states of a black hole has been discussed. Our result shows that some black holes cannot absorb ``energy quantum" under certain conditions from the view of quantization. By quantization of the black hole horizon area, the relation between the number of quanta of area and the microscopic degrees of freedom of the black hole has been shown.
Namely, the relationship between the number of black hole molecules and the energy level of black hole has been established. Thus, the thermodynamic molecular hypothesis of black hole can be understood from the perspective of energy level of black hole, which can advance our knowledge of black holes thermodynamics.
A latent heat of thermodynamical phase transition can be interpreted as
a transition between the number of quanta of area of LBH and the number of quanta of area of SBH in the charged AdS Black Hole. Moreover, based on the Ruppeiner thermodynamic geometry, we have also studied microscopic properties of charged AdS black holes by giving the expression of the Ruppeiner scalar curvature connecting with the number of quanta of area. This provides important insight into the interaction among micromolecules of charged AdS black holes from quantized viewpoint.

  \section{Acknowledgments}

  We would like to thank the National Natural Science Foundation of
  China~(Grant No.11571342) for supporting us on this work.

  \section{References}

 \bibliographystyle{unsrt}
 \bibliography{reference}

\begin{thebibliography}{10}

\bibitem{Hawking1975}
S.~W. Hawking.
\newblock Particle creation by black holes.
\newblock {\em Communications in Mathematical Physics}, 43(3):199--220, Aug
  1975.

\bibitem{PhysRevD.7.2333}
Jacob~D. Bekenstein.
\newblock Black holes and entropy.
\newblock {\em Phys. Rev. D}, 7:2333--2346, Apr 1973.

\bibitem{Bekenstein1972}
J.~D. Bekenstein.
\newblock Black holes and the second law.
\newblock {\em Lettere al Nuovo Cimento (1971-1985)}, 4(15):737--740, Aug 1972.

\bibitem{Hawking1983}
S.~W. Hawking and Don~N. Page.
\newblock Thermodynamics of black holes in anti-de sitter space.
\newblock {\em Communications in Mathematical Physics}, 87(4):577--588, Dec
  1983.

\bibitem{1126-6708-1999-04-024}
Mirjam Cvetic and Steven~S. Gubser.
\newblock Phases of r-charged black holes, spinning branes and strongly coupled
  gauge theories.
\newblock {\em Journal of High Energy Physics}, 1999(04):024, 1999.

\bibitem{HENNEAUX1984415}
Marc Henneaux and Claudio Teitelboim.
\newblock The cosmological constant as a canonical variable.
\newblock {\em Physics Letters B}, 143(4):415 -- 420, 1984.

\bibitem{TEITELBOIM1985293}
Claudio Teitelboim.
\newblock The cosmological constant as a thermodynamic black hole parameter.
\newblock {\em Physics Letters B}, 158(4):293 -- 297, 1985.

\bibitem{0264-9381-28-12-125020}
Brian~P Dolan.
\newblock The cosmological constant and black-hole thermodynamic potentials.
\newblock {\em Classical and Quantum Gravity}, 28(12):125020, 2011.

\bibitem{kubizvnak2012p}
David Kubiz{\v{n}}{\'a}k and Robert~B Mann.
\newblock P-v criticality of charged ads black holes.
\newblock {\em Journal of High Energy Physics}, 2012(7):33, 2012.

\bibitem{Kastor:2009wy}
David Kastor, Sourya Ray, and Jennie Traschen.
\newblock {Enthalpy and the Mechanics of AdS Black Holes}.
\newblock {\em Class. Quant. Grav.}, 26:195011, 2009.

\bibitem{ruppeiner1995riemannian}
George Ruppeiner.
\newblock Riemannian geometry in thermodynamic fluctuation theory.
\newblock {\em Reviews of Modern Physics}, 67(3):605, 1995.

\bibitem{ruppeiner2014thermodynamic}
George Ruppeiner.
\newblock Thermodynamic curvature and black holes.
\newblock In {\em Breaking of Supersymmetry and Ultraviolet Divergences in
  Extended Supergravity}, pages 179--203. Springer, 2014.

\bibitem{ruppeiner2008thermodynamic}
George Ruppeiner.
\newblock Thermodynamic curvature and phase transitions in kerr-newman black
  holes.
\newblock {\em Physical Review D}, 78(2):024016, 2008.

\bibitem{wei2015insight}
Shao-Wen Wei, Yu-Xiao Liu, et~al.
\newblock Insight into the microscopic structure of an ads black hole from a
  thermodynamical phase transition.
\newblock {\em Physical review letters}, 115(11):111302, 2015.

\bibitem{bekenstein1974quantum}
Jacob~D Bekenstein.
\newblock The quantum mass spectrum of the kerr black hole.
\newblock {\em Lettere al Nuovo Cimento (1971-1985)}, 11(9):467--470, 1974.

\bibitem{bekenstein1973black}
Jacob~D Bekenstein.
\newblock Black holes and entropy.
\newblock {\em Physical Review D}, 7(8):2333, 1973.

\bibitem{hod1998bohr}
Shahar Hod.
\newblock Bohr's correspondence principle and the area spectrum of quantum
  black holes.
\newblock {\em Physical Review Letters}, 81(20):4293, 1998.

\bibitem{altamirano2014thermodynamics}
Natacha Altamirano, David Kubiz{\v{n}}{\'a}k, Robert~B Mann, and Zeinab
  Sherkatghanad.
\newblock Thermodynamics of rotating black holes and black rings: phase
  transitions and thermodynamic volume.
\newblock {\em Galaxies}, 2(1):89--159, 2014.

\bibitem{zangeneh2016comment}
M~Kord Zangeneh, A~Dehyadegari, and A~Sheykhi.
\newblock Comment on" insight into the microscopic structure of an ads black
  hole from a thermodynamical phase transition".
\newblock {\em arXiv preprint arXiv:1602.03711}, 2016.

\bibitem{miao2018thermal}
Yan-Gang Miao and Zhen-Ming Xu.
\newblock Thermal molecular potential among micromolecules in charged ads black
  holes.
\newblock {\em Physical Review D}, 98(4):044001, 2018.

\end{thebibliography}
\end{document}